# Temperature dependence of bulk viscosity in water using acoustic spectroscopy
*Journal of Physics: Conference Series*

**M J Holmes[1], N G Parker and M J W Povey**
School of Food Science and Nutrition. University of Leeds, LS2 9JT, UK

E-mail: prcmjh@leeds.ac.uk

**Abstract**. Despite its fundamental role in the dynamics of compressible fluids, bulk viscosity has received little experimental attention and there remains a paucity of measured data. Acoustic spectroscopy provides a robust and accurate approach to measuring this parameter. Working from the Navier-Stokes model of a compressible fluid one can show that the bulk viscosity makes a significant and measurable contribution to the frequency-squared acoustic attenuation. Here we employ this methodology to determine the bulk viscosity of Millipore water over a temperature range of 7 to 50°C. The measured attenuation spectra are consistent with the theoretical predictions, while the bulk viscosity of water is found to be approximately three times larger than its shear counterpart, reinforcing its significance in acoustic propagation. Moreover, our results demonstrate that this technique can be readily and generally applied to fluids to accurately determine their temperature dependent bulk viscosities.

## 1. Introduction

The parameter of bulk or volume viscosity $\mu$ receives little attention, despite being defined in the Navier-Stokes equation for a compressible liquid. It is of fundamental importance in the field of acoustic propagation where it plays the equivalent role in viscosity to the bulk modulus $K$ in solids. Specifically, the total viscosity for fluids $\eta_{total} = \mu + 4\eta/3$ is related to the longitudinal modulus $M = K + 4G/3$ in solids, where $G$ is the shear modulus and $\eta$ is shear viscosity, through the time dependence relation $i\omega\eta_{total} \rightarrow M$. In the absence of scattering, the classical theory of acoustic attenuation predicts a frequency-squared dependence on the viscosity. Shear viscosity is routinely measured using commercial rheological instrumentation, and bulk viscosity may be determined from a measurement of the frequency dependent attenuation, the shear viscosity and the ratio of specific heats for the fluid. However, very few measurements of this rheological parameter have been made despite its fundamental importance in the dynamics of compressible fluids.

The general description of the equations of motion for sound propagation in a compressible fluid is supplied by the Navier-Stokes and continuity equations [1, 2] in which the parameter of bulk viscosity is a natural consequence. Additionally, energy and thermodynamic equations are introduced [3-5] which describe the energy in the system using the state variables temperature, pressure and density. Considering only first order terms in time and assuming periodic state solutions of the form $\exp(i\omega t)$ enables the equations to be simplified by replacing partial time derivatives with the factor $-i\omega$ (where $\omega=2\pi f$ is the angular frequency, $i$ is the imaginary number and $f$ is the frequency of the acoustic wave) to yield the following equations (1) and (2),

$$\omega^2 \mathbf{v} + \left(\frac{v^2}{\gamma} - i\omega \frac{N\eta}{\rho}\right) \nabla(\nabla \cdot \mathbf{v}) + i\omega \frac{\eta}{\rho} \nabla \times \nabla \times \mathbf{v} = \frac{i\omega \beta v^2}{\gamma} \nabla T, \qquad N = \left(\frac{4}{3} + \frac{\mu}{\eta}\right) \qquad (1)$$

$$\gamma \frac{\tau}{\rho C_p} \nabla^2 T + i\omega T = \frac{(\gamma - 1)}{\beta} \nabla \cdot \mathbf{v} \ . \qquad (2)$$

Here $\mathbf{v}$ is the velocity vector of the fluid, $T$ is the temperature, $\gamma$ is the ratio of specific heats, $\tau$ is the thermal conductivity, $\rho$ is the density, $C_p$ is the specific heat at constant pressure, $\beta$ is the bulk compressibility, $\eta$ is the shear viscosity, $\mu$ is the bulk viscosity and $v$ is the velocity of sound. In equation (1) the physical origins of the bulk viscosity is clear: it acts as a dissipative coefficient to dilatational-compressional motion of the fluid (described by the $\nabla\cdot\mathbf{v}$ term). Thus one can expect the bulk viscosity to play a role in fluid dynamics wherever such motions exist, notably in acoustic propagation and shock waves.

---
[1] To whom any correspondence should be addressed.

Equation (1) can be further simplified by representing the velocity vector **v** in terms of scalar potentials which describe the longitudinal compressional field $\varphi$ and the thermal field $\psi$ together with a transverse shear vector potential **A** in the form, $\mathbf{v} = -\nabla(\varphi + \psi) + \nabla \times \mathbf{A}$. This allows equations (1) and (2) to be combined into a biharmonic type equation which may be then decoupled into independent Helmholtz equations for each given field as shown in equations (3) with respective acoustic, thermal and shear wavenumbers $K$, $L$ and $M$ (see [6-9] for full details),

$$(\nabla^2 + K^2)\varphi = 0 \qquad (\nabla^2 + L^2)\psi = 0 \qquad (\nabla^2 + M^2)\mathbf{A} = 0. \qquad (3)$$

In particular, the wavenumbers $K$, $L$ and $M$ are given by,

$$K = \frac{\omega}{v} + i\alpha = \frac{\omega}{v} + i\frac{\eta\omega^2}{2\rho v^3}\left[\frac{4}{3} + \frac{\mu}{\eta} + \frac{(\gamma-1)\tau}{\eta C_p}\right], \qquad (4)$$

and,

$$L = \left(\frac{\omega\rho C_p}{2\tau}\right)^{1/2}(1+i), \qquad M = \left(\frac{\omega\rho}{2\eta}\right)^{1/2}(1+i). \qquad (5)$$

The dissipative terms in the wavenumbers are characterised by imaginary components, with the imaginary term in equation (4) parameterising the attenuation of the longitudinal acoustic wave. Here the three terms forming from attenuation arise from three distinct sources: shear viscosity, bulk viscosity and thermal conductivity. Importantly, we see from equation (4) that this acoustic attenuation scales as the square of the frequency. Since this prediction is based upon the Navier-Stokes equation, which by definition describes a Newtonian fluid, a deviation from frequency-squared attenuation spectrum offers a signature of non-Newtonian fluidity. Thus acoustic attenuation measurements provide an independent test of Newtonian behaviour to complement the well-established techniques based on shear rheometry.

From equation (4) we see that the total acoustic attenuation features a significant contribution from the bulk viscosity, highlighting the importance of considering the bulk viscosity when dealing in acoustic propagation. Equation (4) provides the theoretical and experimental route in calculating the bulk viscosity: by measuring the attenuation and sound speed, and assuming other parameters are either known or independently measurable, then one may calculate the bulk viscosity through the equation,

$$\mu = \left[\frac{2\alpha\rho v^3}{\omega^2} + \frac{4\eta}{3} + \frac{(\gamma-1)\tau}{C_p}\right]. \qquad (6)$$

Dukhin and Goetz [10] compared three techniques for bulk viscosity measurement: Brillouin spectroscopy, laser transient grating spectroscopy and acoustic spectroscopy. They concluded that acoustic spectroscopy is best placed and has the advantage of theoretical verification. They then proceeded to determine the bulk viscosity of water and various other fluids at a fixed temperature of 25°C. In this paper, we build upon the work conducted by Dukhin and Goetz [10] by extending the experimental measurements of bulk viscosity to the temperature domain. Specifically, we measure the bulk viscosity for Millipore water and reveal its temperature dependence over the range 7-50°C. In the next section we shall describe the measurement procedure for accurately determining the attenuation spectrum, and in Section 3 we will present and discuss the validation and results of our measurements, notably the temperature-dependence of the bulk viscosity of water. This work is novel in that a single study presents the bulk viscosity over several temperatures and, additionally, determines a functional fit to the temperature dependence. This approach may then readily be pursued in a systematic way to determine temperature dependent bulk viscosities of other fluids.

2. **Measurement Techniques and Instruments**

Acoustic attenuation measurements were carried out using an Ultrasizer MSV by Malvern Ltd [11] (see Figure 1), an instrument which performs acoustic spectroscopy of liquids and emulsions in the frequency range 1 – 100MHz. In brief, the instrument operates in transmission mode where one transducer emits sound into the sample while an opposing transducer detects the sound waves and generates an associated voltage signal. Two such pairs, one operating at low frequency and one at high frequency, are employed. Due to their broad-band response, the acoustic frequency of each pair can be swept over a wide range. The attenuation is measured from the decay of the received voltage with propagation distance through the sample. The diffraction of the beam, which also causes variations in the received voltage, is calibrated automatically and its effects factored out. Furthermore, by comparing the received signal at two propagation distances the non-trivial response of the transducer and electronic circuitry and any other systematic errors can also be factored out. Hence, an absolute measure of the acoustic attenuation is determined. Finally, the transducer separation is calculated automatically for a given fluid so as to exploit the full dynamic range of the measurement system and maximise the signal-to-noise ratio.

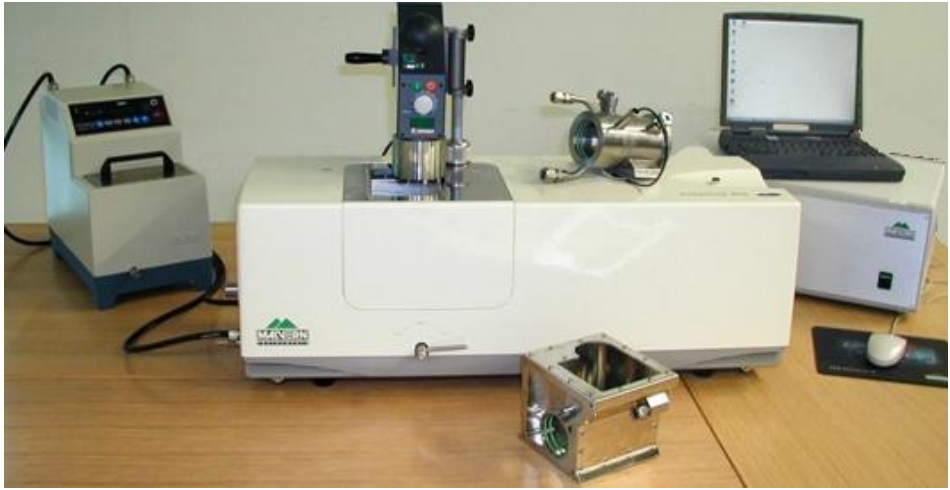

**Figure 1**. The Malvern Ultrasizer™ [11] with its associated temperature control unit. The sample vessel is shown foremost in the picture. This requires 500ml of the sample under test. During the measurement process, the sample is agitated with a stirrer to reduce thermal fluctuations. The Ultrasizer can perform attenuation measurements in the frequency range 1-100MHz.

The Ultrasizer requires 500ml of sample and so, although offering an effective measurement system, is not suited for limited sample volumes. The Ultrasizer is connected to an external Huber Ministat temperature control unit which enables the sample temperature to be varied across a range of 0 – 50°C. During the measurement process the sample under test is agitated by a stirrer whose angular speed can be varied in the range 0-500 rpm. The agitator is employed to reduce thermal fluctuations in the sample (for emulsions it also suppresses sedimentation and creaming). This combination maintains the sample at the selected temperature to within ± 0.5°C, as measured by the internal thermometer of the Ultrasizer, and so provides a stable temperature to determine the associated acoustic attenuation. Note that the acoustic power levels are of the order of milliwatts and so the associated heating of the sample is negligible.

| Temperature | Density | Shear viscosity | Thermal conductivity | Isobaric heat capacity | Isochoric heat capacity | Ratio of specific heats | Speed of sound |
|---|---|---|---|---|---|---|---|
| $T$ | $\rho$ | $\eta$ | $\tau$ | $C_P$ | $C_V$ | $\gamma$ | $v$ |
| °C | kg m$^{-3}$ | Pa s | W m$^{-1}$ K$^{-1}$ | J kg$^{-1}$ m$^{-3}$ | J kg$^{-1}$ m$^{-3}$ | | m s$^{-1}$ |
| 7 | 999.81 | 1.43E-03 | 0.5747 | 4200.6 | 4199.2 | 1.0003 | 1434.92 |
| 10 | 999.70 | 1.31E-03 | 0.5800 | 4195.6 | 4190.4 | 1.0012 | 1447.29 |
| 15 | 998.97 | 1.14E-03 | 0.5900 | 4188.7 | 4174.2 | 1.0035 | 1465.96 |
| 25 | 997.00 | 8.88E-04 | 0.6075 | 4181.5 | 4137.5 | 1.0106 | 1496.73 |
| 40 | 992.22 | 6.53E-04 | 0.6305 | 4177.6 | 4073.3 | 1.0256 | 1528.89 |
| 50 | 988.03 | 5.47E-04 | 0.6435 | 4181.4 | 4026.1 | 1.0386 | 1542.57 |

**Table 1.** Parameter values for water at the 6 temperatures under investigation. Density, shear viscosity and thermal conductivity are derived from [12], and the speed of sound is taken from [13]. The isobaric and isochoric heat capacities were derived from the IAPWS formulated data in [14], and the ratio of specific heats determined via $\gamma = C_P/C_V$. Where data was not present for our specific temperatures we employed polynomial [12, 13] and spline [14] fitting to determine the required values.

In this work we consider 0.22$\mu$m Millipore water as the sample. We restrict ourselves to the temperature range of 7 - 50°C. The lower limit is set to avoid microbubble formation at low temperature which can significantly affect ultrasonic measurements and potentially introduce erroneous attenuation measurements (see [13] for example). The upper limit is set by the operating conditions of the instrument. The instrument implements automated processes to conduct measurements at 50 different frequencies selected within the range 10 – 100 MHz. NB we set the lower bound to be 10 MHz rather than the instrument limit of 1 MHz since attenuations at low frequencies can be minimal and difficult to measure.

To determine the bulk viscosity from equation (6) and measured attenuation values, it is required to know all of the other thermodynamic parameters appearing in the equation. We retrieve these values from experimental data in references [12-14] and they are tabulated in Table 1.

At each selected frequency of the acoustic spectra, 30 repeat attenuation measurements were made and, from these, an average attenuation value was calculated. We have verified that these 30 data values were symmetric about the mean value and could be fitted to a normal distribution. Finally, using the mean attenuation values across the chosen temperatures we calculated the mean bulk viscosity according to equation (6). Both average attenuation values and their standard error can then be quoted (as performed in Table 2).

## 3. Results and Discussion

Our first task was to verify that the measured attenuation was not affected by the agitation speed. In Figure 2 we present (on a logarithmic scale) the attenuation spectrum of Millipore water at 25°C for agitator speeds of 150, 300 and 450 rpm. As can be seen, no discernable difference exists and so we subsequently measured all other samples using the intermediate speed of 300 rpm.

For the 300rpm results in Figure 2 we have fit them with a single polynomial in frequency of the form $f^{\delta}$. We find that an excellent fit ($R^2$=1) is provided by an exponent of $\delta$ =2.0012. Indeed, we observe that all of our spectra give an exponent within a similar proximity to 2. We are thus well justified in fitting to the frequency-squared prediction of equation (4). Note that the attenuation arising from thermal conductivity is found to give a negligible contribution to the overall attenuation. According to equation (4) this contribution is proportional to ($\gamma$-1). For water (and indeed for most liquids), the ratio of specific heats remains very close to unity due to its low compressibility (e.g., see values in table 1) and thus the thermal term in equation (4) remains small compared to the viscous terms.

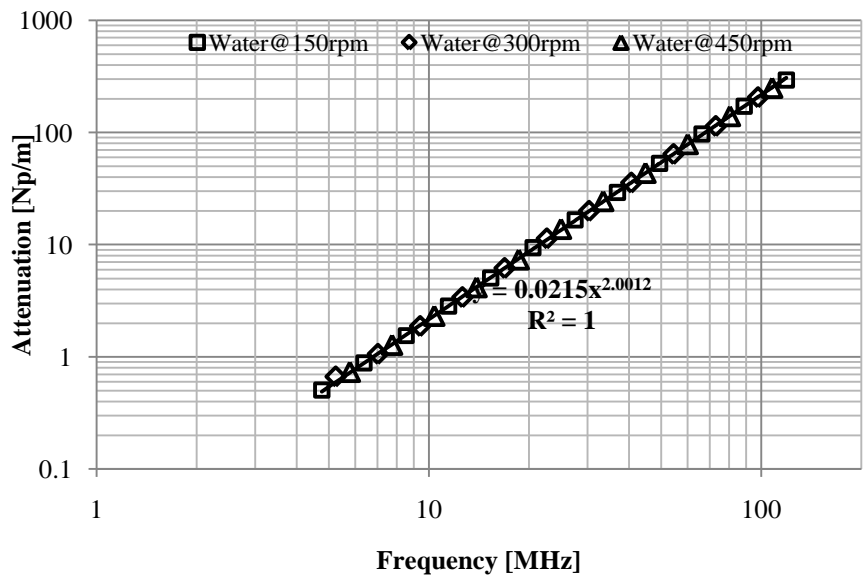

**Figure 2**. Attenuation-frequency plot for 0.22$\mu$m Millipore water at 25°C with agitation speeds 150, 300 and 450 rpm. As can be seen no appreciable difference was detected. Consequently all measurements were made using the intermediate agitation of 300rpm. A polynomial best fit (for the 300 rpm case) is illustrated.

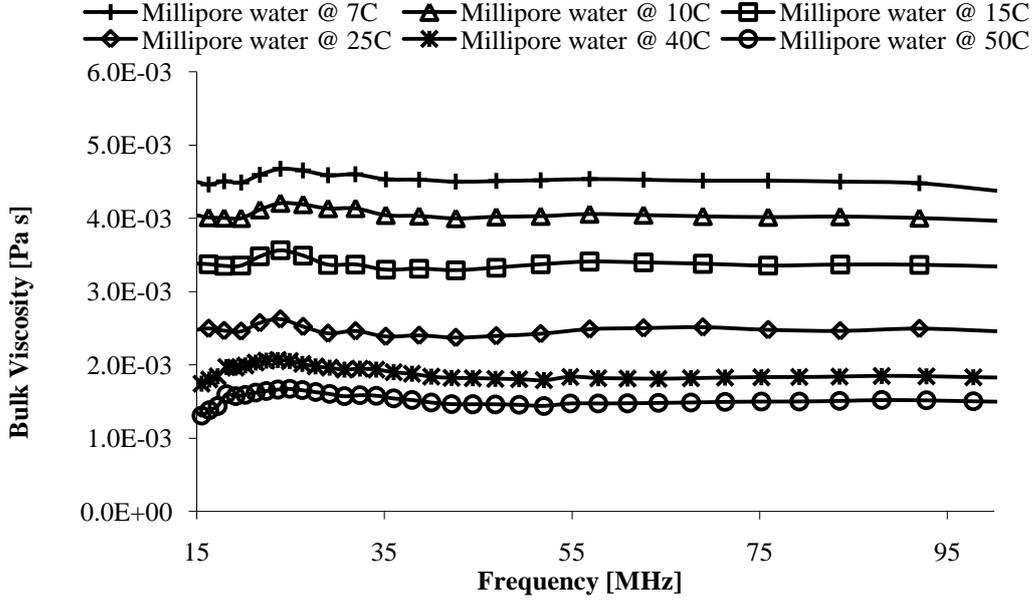

**Figure 3**. Bulk viscosity [Pa.s] for 0.22$\mu$m Millipore water against frequency [MHz] for the temperatures 7, 10, 15, 25, 40 and 50°C. We observe the bulk viscosity to be approximately constant with frequency, as expected for a Newtonian fluid.

| Temperature | Bulk Viscosity | Standard error | Attenuation | Bulk viscosity/ shear viscosity |
|---|---|---|---|---|
| $T$ | $\mu$ | | $\alpha/f^2$ | $\mu/\eta$ |
| °C | Pa s | Pa s | Np m$^{-1}$ MHz$^{-2}$ | |
| 7 | 4.50E-03 | 1.20E-05 | 4.28E-02 | 3.15 |
| 10 | 4.03E-03 | 1.17E-05 | 3.76E-02 | 3.08 |
| 15 | 3.38E-03 | 1.10E-05 | 3.07E-02 | 2.96 |
| 25 | 2.47E-03 | 1.08E-05 | 2.16E-02 | 2.78 |
| 40 | 1.84E-03 | 2.70E-05 | 1.51E-02 | 2.82 |
| 50 | 1.48E-03 | 2.76E-05 | 1.21E-02 | 2.71 |

**Table 2.** The measured bulk viscosity, bulk viscosity standard error, attenuation, and bulk-to-shear ratio in Millipore water for six selected temperatures. For a selected temperature, at each frequency, mean attenuation values are computed by averaging over 30 measured attenuation values. A total mean bulk viscosity and its standard error are then computed by averaging over all frequencies. Similarly, the attenuation value $\alpha/f^2$ presented is computed in the same fashion. For comparison the ratio of the bulk to shear viscosity is also shown.

In Figure 3 we have plotted the bulk viscosity, as calculated using the measured attenuation $\alpha$ and equation (6), as a function of frequency. Each line corresponds to a different temperature. The bulk viscosity remains approximately constant with frequency, as expected for a Newtonian fluid.

In table 2 we show the calculated values of the bulk viscosity and attenuation at each temperature considered. For the purposes of comparison we also tabulated the ratio of the bulk viscosity to the shear viscosity. The bulk viscosity is approximately three times its shear counterpart throughout the temperature range considered.

In [10] Dukhin and Goetz reported a bulk viscosity value of 2.43E-03 Pa.s at 25°C for distilled water, which is in very good agreement with our value of 2.469E-03 [2.458E-03, 2.479E-03].

Similarly, they reported a ratio of bulk to shear viscosity as 2.73 whereas we obtained a value of 2.78. Another example cited in [10] is the result obtained by Litovitz and Davis [4] who reported a value of 3.09E-03 for the bulk viscosity of water at 15°C and viscosity ratio of 2.81, whereas in our case we record a value of 3.38E-03, with ratio 2.96.

In Figure 4 we have plotted the frequency-averaged bulk viscosity as a function of temperature. For comparison, we plot the shear viscosity of pure water obtained from the literature [12]. The bulk viscosity is approximately 3 times larger than its shear counterpart throughout the temperature range, and both decrease with temperature in a similar monotonic form. We also plot the bulk viscosities determined from several other studies [4, 10, 15]. References [4] and [10] report the bulk viscosity at a single temperature. However, Xu et al [16] used Brillioun scattering to determine the bulk viscosity of water over a similar temperature range to this work. As can be seen, our results concur well with those obtained elsewhere.

A common empirical model for the temperature-dependence of shear viscosity is that of an exponential decay,

$$\mu = A_1 \exp(-A_2 T), \qquad (7)$$

($A_1$ and $A_2$ empirical constants) as first observed by Reynolds in 1886 [15]. Our bulk viscosity data closely follows such a model well with coefficients $A_1$=5.091712E-03 and $A_2$=2.545425E-02 ($R^2$=0.986). However, our data is most closely fitted by a cubic expression of the form,

$$\mu = B_0 + B_1 T + B_2 T^2 + B_3 T^3 \qquad (8)$$

with coefficients $B_0$ = 5.94068E-03, $B_1$ = - 2.37073E-04, $B_2$ = 4.94789E-06 and $B_3$ = -3.97502E-08 ($R^2$ =0. 99991) .

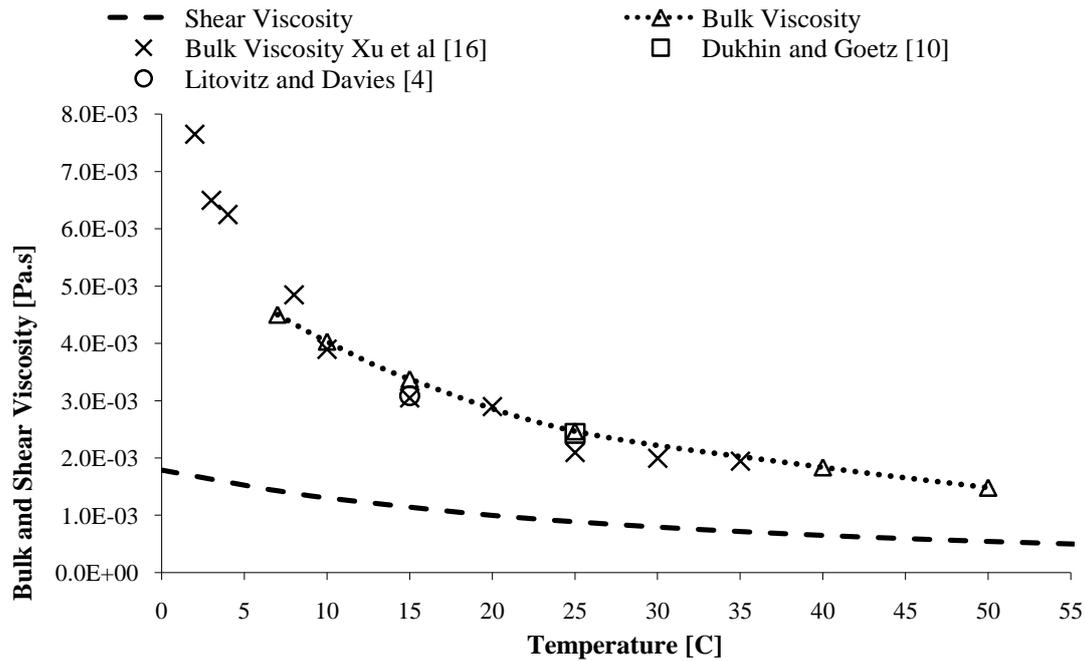

**Figure 4**. Bulk viscosity (··△··) [Pa s] for 0.22$\mu$m Millipore water against temperature [°C] using acoustic spectroscopy. Shear viscosity (- - -) taken from [12].Bulk viscosity data (×) by Xu et al [16] using Brillioun scattering, (o) Litovitz and Davies [4] and (□) Dukhin and Goetz [10].

**4. Conclusions**
We have successfully employed acoustic spectroscopy to determine the bulk viscosity of 0.22$\mu$m Millipore water across the temperature range 7-50°C. We have shown results consistent with the classical theory of acoustic attenuation exhibits a frequency squared dependence on the viscosity. The data obtained has shown that the bulk viscosity of water decays with temperature, and we specify exponential and polynomial fits of high $R^2$ value. Our results are in good agreement with other studies that have measured bulk viscosity using both the same and different techniques. Our work further demonstrates that acoustic spectroscopy provides a robust, repeatable and accurate approach to determining the bulk viscosity. The ratio of the bulk viscosity to shear viscosity is equal to approximately three throughout this temperature range, indicating that it cannot be ignored when dealing with thermal and fluid systems. The ratio decreases slightly with temperature, suggesting a convergence at higher temperatures, although further study is required to probe this. Our experimental measurements extend the limited bank of fluid bulk viscosity measurements that have been made to date. Such continued exploration of this fundamental fluid parameter is important to develop new fluid characterisation methods and provide theoretical studies with much needed data [17].

**Acknowledgement**
The authors would like to thank Dahe Liu for supplying the bulk viscosity data given in [16].